\documentclass{article}
\usepackage{natbib}

\usepackage{microtype}
\usepackage{graphicx}
\usepackage{subfigure}
\usepackage{booktabs} 

\usepackage{hyperref}

\usepackage[accepted]{icml2025}

\usepackage{amsmath}
\usepackage{amssymb}
\usepackage{mathtools}
\usepackage{amsthm}

\usepackage[capitalize,noabbrev]{cleveref}

\theoremstyle{plain}

\theoremstyle{definition}

\theoremstyle{remark}

\usepackage[textsize=tiny]{todonotes}

\icmltitlerunning{TMA-Adaptive FP8 Grouped GEMM}

\begin{document}

\twocolumn[
\icmltitle{TMA-Adaptive FP8 Grouped GEMM: Eliminating Padding Requirements in Low-Precision Training and Inference on Hopper}



\icmlsetsymbol{equal}{*}

\begin{icmlauthorlist}
\icmlauthor{Zhongling SU}{comp}
\icmlauthor{Rong FU}{comp}
\icmlauthor{Weihan CAO}{comp}
\icmlauthor{Jianfei GAO}{comp}
\icmlauthor{Minxi JIN}{comp}
\icmlauthor{Zhilin PEI}{comp}
\icmlauthor{Hui WANG}{comp}
\end{icmlauthorlist}

\icmlaffiliation{comp}{Shanghai Artificial Intelligence Laboratory, Shanghai, China}
\noindent
{\centering \textsuperscript{1} \textbf{Shanghai Artificial Intelligence Laboratory}, \textbf{Shanghai, China} \par}

\icmlkeywords{FP8 Grouped GEMM, Low-Precision Computation, Tensor Memory Accelerator, Padding-Free GEMM, High-Performance Computing, MOE}

\vskip 0.3in
]




\begin{abstract}

Current FP8 grouped GEMM implementations require padding each group to a fixed alignment (e.g., 128), incurring memory and computational overhead. We propose \textit{TMA-Adaptive FP8 Grouped GEMM}, which eliminates padding by dynamically adapting to variable group dimensions via (1) a TMA descriptor pool with $\log_2(block_M)$ preconfigured descriptors to handle all residual row cases through dynamic runtime selection and dual-phase load-store operations, achieving comprehensive coverage with minimal overhead, and (2) TMA-alignment-aware management to satisfy 16-byte global memory alignment and 128-byte shared memory alignment. Experiments demonstrate 1.7\% to 20.4\% speed up with up to 23.8\% memory reduction compared to padding operation plus state-of-the-art FP8 grouped GEMM, while maintaining full numerical equivalence for valid data. The source code is publicly available at an anonymous repository: \url{https://github.com/sukoncon/TMA-Adaptive-FP8-Grouped-GEMM}.


\end{abstract}

\section{Introduction and Related Work}
\label{intro}

To accelerate computation and reduce memory consumption, the exploration of low-precision computation has gained momentum. In 2018, \cite{half_training} pioneered the training of deep neural networks using half-precision floating-point numbers. With subsequent hardware advancements, lower precision techniques continued to evolve, as demonstrated by \cite{int8_training} which proposed SwitchBack - an int8 quantized linear layer for training. The stability and computational efficiency of 8-bit training/inference for large models were ultimately validated by \cite{deepseekv3}.

Modern GPU architectures Hopper and beyond provide support for FP8 tensor core. In fully utilize GPU capabilities, various optimization techniques have been developed \cite{deepgemm,flashattention3}, including warp specialization, Tensor Memory Accelerator (TMA) \cite{TMA}, and warp group matrix multiply-accumulate, collectively maximizing memory bandwidth utilization and arithmetic intensity for foundation model workloads.

These low-precision techniques find particularly important applications in specialized architectures. Grouped General Matrix Multiplication (Grouped GEMM) operations play a critical role in Mixture-of-Experts (MoE) architectures, where dynamic group sizes arise from variable sequence lengths selected via top-k routing.

In Grouped GEMM computations, padding is essential due to hardware constraints when handling variable dimensions (e.g., dynamically routed sequence lengths in MOE). The necessity arises from two fundamental architectural limitations: First, the static configuration of TMA descriptors during host initialization \textbf{prevents dynamic adjustment to varying row dimensions across different groups}. Second, the Hopper architecture's TMA imposes strict alignment requirements - \textbf{16-byte for global memory and 128-byte for shared memory} for multi-dimensional bulk tensor asynchronous copies \cite{cudaProgrammingGuide}. 

In this work, we present TMA-Adaptive FP8 Grouped GEMM, a hardware-compliant optimization framework for FP8 matrix multiplication that eliminates padding overhead while strictly satisfying TMA alignment constraints. Our solution introduces two key innovations:

First, a TMA descriptor pool that predefines descriptors for all possible residual rows (the remaining rows when the group row count is divided by $\mathtt{block_M}$), followed by dynamic runtime descriptor selection and dual-phase load-store operations. This approach \textbf{achieves coverage of all possible residual row cases with only $\log_2(block_M)$ preconfigured descriptors}. 

Second, we propose an TMA-alignment-aware memory access scheme that guarantees compliance through two mechanisms: over-fetching to maintain 16-byte global memory alignment, and constraining $\mathtt{block_N}$ to multiples of 64 elements to ensure 128-byte shared memory alignment boundaries.

In the following sections, we first present our methodology in Section~\ref{sec:methodology}, followed by experimental results and analysis in Section~\ref{experiment}. Additional details are provided in the Appendix.


\section{Methodology}
\label{sec:methodology}

\begin{figure*}[ht]
\centering
\includegraphics[width=\textwidth]{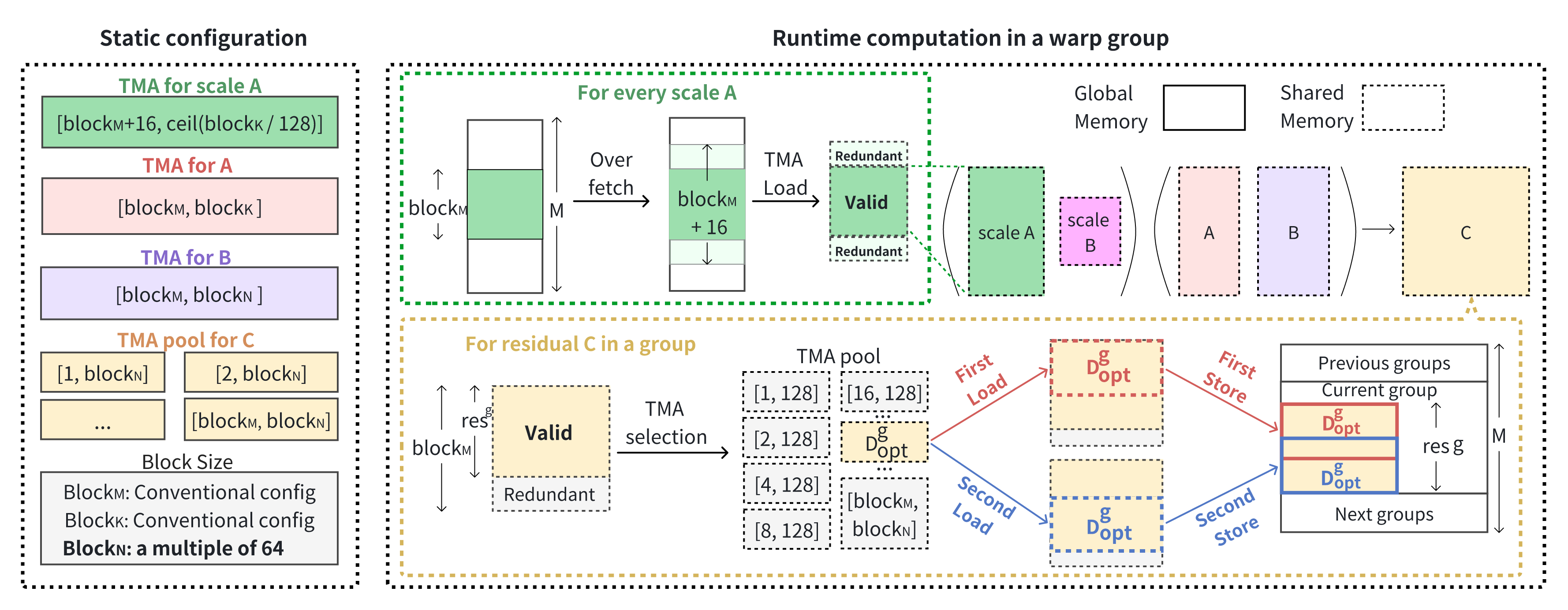}
\vspace{-0.3in}
\caption{The framework of TMA-Adaptive FP8 Grouped GEMM. The left panel shows the static configuration featuring our proposed TMA pool for C and block size constraint for $N$. The right panel illustrates the runtime computation within a warp group. Key innovations are highlighted: (1) global memory prefetch for scaled blocks of matrix A (green), and (2) Dynamic Descriptor Selection with Two-Phase Load-Store for residual elements in matrix C (yellow).}
\label{framework}
\end{figure*}



We present an optimization framework (Figure~\ref{framework}) featuring two key innovations. In this section, we first introduce the fundamental concepts of grouped GEMM, then provide detailed descriptions of the two major innovations: dynamic descriptor selection with two-phase load-store in Section~\ref{subsec:tma_pool} (corresponding to the yellow blocks in Figure~\ref{framework}) and TMA-Alignment-Aware memory management in Section~\ref{subsec:overfetch} (addressing the green blocks and $block_N$ configuration in Figure~\ref{framework}).




\subsection{Notation and Preliminaries for Grouped GEMM}
\label{subsec:group_gemm}

\begin{table}[htbp]
\vspace{-0.3in}
\small
\centering
\setlength{\tabcolsep}{2pt} 
\caption{Key notation for grouped matrix multiplication}
\label{tab:notation}
\begin{tabular}{@{}cll@{}}
\toprule
\multicolumn{1}{c}{\textbf{Symbol}} & \textbf{Description} & \textbf{Dimensions} \\ 
\midrule
$g$ & Group index & Scalar \\
$M^g$ & Variable row dimension of group $g$ & Scalar \\
$N$ & Output column dimension & Scalar \\
$K$ & Hidden dimension & Scalar \\
$\mathbf{A}^g$ & Left matrix for group $g$ & $\mathbb{R}^{M^g \times K}$ \\
$\mathbf{B}^g$ & Right matrix for group $g$ & $\mathbb{R}^{K \times N}$ \\
$\mathbf{C}^g$ & Output matrix for group $g$ & $\mathbb{R}^{M^g \times N}$ \\
$\mathbf{S}_A^g$ & Per-tile scale for $\mathbf{A}^g$ & $\mathbb{R}^{M^g \times \lceil K/128 \rceil}$ \\
$\mathbf{S}_B^g$ & Per-block scale for $\mathbf{B}^g$ & $\mathbb{R}^{\lceil K/128 \rceil \times \lceil N/128 \rceil}$ \\
\bottomrule
\end{tabular}
\end{table}
\vspace{-0.15in}

Table \ref{tab:notation} summarizes the key notation used in Grouped GEMM with specialized quantization schemes. Following \cite{deepseekv3}, we employ 1x128 tiled scaling for $A$ and 128x128 blocked scaling for $B$.

In the context of mixture-of-experts (MoE) architectures for LLMs, the dimension $M_g$ denotes the \textit{dynamically routed sequence length} assigned to expert group $g$. The static parameters $K$ (hidden dimension) and $N$ (output dimension) remain constant across all groups. 

The CUDA kernel configurations $\mathit{block_M}$, $\mathit{block_N}$, and $\mathit{block_K}$ represent fixed tiling sizes for computational optimization.


Among all operands involved in Grouped GEMM, particular attention must be devoted to memory alignment when \textbf{loading $S_A^g$ from global memory} and \textbf{loading $C^g$ from shared memory}. A detailed analysis of memory alignment is provided in Appendix \ref{alignment_analysis}. 

\subsection{Dynamic Descriptor Selection with Two-Phase Load-Store}
\label{subsec:tma_pool}

Our primary technical contribution is a hardware-aware TMA descriptor design that combines static configuration compliance with dynamic runtime adaptation, featuring a novel two-phase load-store mechanism for \textbf{residual blocks of output matrix}.


The methodology comprises three stages:

\textbf{Descriptor Pool Predefinition:} During kernel initialization, we construct a descriptor pool (see static configuration of Figure~\ref{framework}) derived from the block dimension $block_M$:
    \begin{equation}
        \mathcal{D}_{pool} = \left\{ [2^i, block_N] \;\big|\; i \in \mathbb{N},\ 0 \leq i \leq \lfloor \log_2(block_M) \rfloor \right\}
    \end{equation}
    
\textbf{Runtime Selection:} At runtime, for each group $g$ with residual rows $res^g = M^g \mod block_M$, we determine the optimal descriptor:
\begin{equation}
    \mathcal{D}_{opt}^g = [2^{\lfloor \log_2(res^g) \rfloor}, block_N]
\end{equation}

\textbf{Two-Phase Load-Store:} The mechanism employs two coordinated TMA operations: (a) Shared memory rows $\left[0, 2^{\lfloor \log_2(res^g) \rfloor}-1\right]$ to global memory rows $\left[M^g - res^g, M^g - res^g + 2^{\lfloor \log_2(res^g) \rfloor}-1\right]$, and (b) Shared memory rows $\left[res^g - 2^{\lfloor \log_2(res^g) \rfloor}, res^g-1\right]$ to global memory rows $\left[M^g - 2^{\lfloor \log_2(res^g) \rfloor}, M^g-1\right]$

This design implements a safe overlapping write strategy by intentionally writing only a small portion of identical data in the overlapping memory regions, ensuring that: (1) preservation of all valid results, (2) strict boundary compliance, (3) logarithmic descriptor scalability ($\lfloor \log_2(block_M) \rfloor$ TMAs for $block_M$ residuals), and (4) two TMA operations per residual block regardless of its size. See Appendix~\ref{TMA_example} for details.

\subsection{TMA-Alignment-Aware Memory Management} 
\label{subsec:overfetch}

Section \ref{subsec:tma_pool} demonstrates that the second TMA load operation starts at row $res^g - 2^{\lfloor \log_2(res^g) \rfloor}$, potentially causing misalignment issues. To guarantee 128-byte alignment regardless of the starting row position, we enforce $block_N$ to be a multiple of 64.

The per-row offset of $4\lceil K/128\rceil$ bytes in global memory for $S_A^g$ may cause misalignment issues. We address this misalignment through boundary-aligned prefetching, where the starting address is adjusted to meet 16-byte alignment requirements. The number of prefetched rows from previous data $row^g_{\text{prev}}$ is determined by:

\vspace{-0.1in}
\begin{equation}
 \min\left\{\left[Addr_{SA}^g - 4row^g_{\text{prev}}\left\lceil\frac{K}{128}\right\rceil \right] \bmod 16 = 0 \right\}
\end{equation}

where $Addr_{SA}^g$ denotes the base address of vector group $S_A^g$ and the coefficient 4 reflects \texttt{float} datatype storage. Subsequent rows $row^g_{\text{next}} = 16 + block_M - row^g_{\text{prev}}$ complete the prefetching window, with TMA descriptors configured for [$16+block_M$, $\lceil K/128 \rceil$] dimensions. Only the central $block_M$ rows (valid green block in Figure~\ref{framework}) participate in computation.


\section{Experiment}
\label{experiment}
The experimental evaluation was performed on an NVIDIA H800 GPU accelerated computing platform using PyTorch 2.6.0 and CUDA 12.6. Our baseline implementation integrates explicit input padding with DeepGEMM, which currently represents the state-of-the-art high-performance FP8 GEMM implementation, against which we compare our optimized approach across three critical metrics: computational acceleration, numerical equivalence, and memory efficiency.

\subsection{Experimental Setup}
The parameter space was designed to reflect practical configurations in modern large language model architectures, with tensor dimensions spanning $N, K \in \{3072, 4096, 5120, 6144, 7168, 8192\}$, group counts in $\{4, 8, 16, 32\}$, and sequence lengths after top-k routing $M \in \{8192, 16384, 32768, 65536\}$ with each group dimension $M^g$ being randomly generated (see Appendix \ref{M_generation}).

We implemented the padding operation for matrices $A$ and $S_A$ using a custom high-performance kernel written in Triton. This kernel achieves approximately 2000 GB/s DRAM bandwidth (2173 GB/s maximum), ensuring the validity of our comparative results against the baseline. 
\subsection{Performance Analysis}

\begin{figure}[ht]
\centering
\includegraphics[width=\columnwidth]{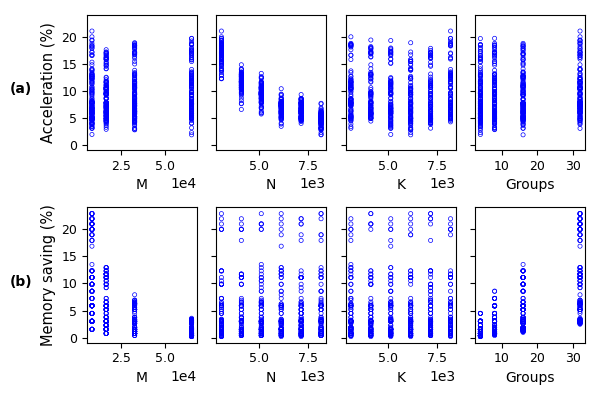}
\vspace{-0.1in}
\caption{Performance comparison between optimized and baseline implementations showing (a) computational acceleration ratios and (b) relative memory savings across varying M, N, K and group counts.}
\label{acc_mem}
\end{figure}

Figure~\ref{acc_mem} demonstrates computational acceleration ratios and memory savings across configurations. 

The acceleration (Figure~\ref{acc_mem}a) ranges from 1.7\% to 20.4\%, exhibiting a weak positive correlation with sequence length $M (r = 0.09)$ and group count $(r = 0.096)$, while showing a strong negative correlation with parameter $N (r = - 0.899)$. The observed acceleration stems from eliminating padding-induced memory transactions, while the dominant anti-correlation with N reflects linear scaling of residual write-back costs.

Memory savings ((b) in Figure~\ref{acc_mem})  demonstrate strong inverse correlation with sequence length $M$ accompanied by near-linear scaling with group count, achieving maximum observed savings of 23.8\% when operating at $M=8192$ with 32 groups. This fundamental relationship stems from geometric padding characteristics in grouped tensor operations - smaller $M$ values combined with higher group counts yeild more padding data.


Numerical equivalence was verified through comparison of these test cases after removing zero-padded elements from baseline outputs, demonstrating bitwise identical results between our method and baseline implementations. This zero-error guarantee stems from our two-phase load and store: the final write-back operation strictly preserves only the valid data region. 

\vspace{-0.1in}

\section{Conclusion}

This paper presents TMA-Adaptive FP8 Grouped GEMM, a hardware-aligned solution that eliminates the requirement for padding groups to fixed alignment multiples (e.g., 128 elements) in low-precision Grouped GEMM. 

By resolving key hardware-alignment conflicts in variable-length scenarios—specifically, static TMA descriptor limitations and memory boundary misalignments—our approach simultaneously enhances memory efficiency and computational throughput without degrading numerical accuracy. 
 
Although our method constrains $block_N$ to multiples of 64 (e.g., 64, 128, 192), these values represent optimal or near-optimal configurations for the N dimension in practice. Experimental validation demonstrates significant improvements, including up to 23.8\% reduction in memory allocation overhead and 1.7–20.4\% end-to-end speedup compared to state-of-the-art padding implementations, while numerical verification confirms bitwise equivalence, preserving accuracy.

By fundamentally eliminating padding operations, our solution establishes an innovative paradigm for Grouped GEMM, delivering superior computational efficiency with reduced memory footprint and seamless plug-and-play compatibility for dynamic routing without kernel modifications beyond Grouped GEMM.

Crucially, this work directly enhances Mixture of Experts (MoE) Large Language Models: it accelerates inference and training by minimizing latency and memory overhead during dynamic expert routing, thereby facilitating scalable high-performance MoE LLM deployments.

\section*{Acknowledgements}
Project supported by Shanghai Municipal Science and Technology Major Project.

\nocite{langley00}

\bibliography{main_paper}

\begin{thebibliography}{7}
\providecommand{\natexlab}[1]{#1}
\providecommand{\url}[1]{\texttt{#1}}
\expandafter\ifx\csname urlstyle\endcsname\relax
  \providecommand{\doi}[1]{doi: #1}\else
  \providecommand{\doi}{doi: \begingroup \urlstyle{rm}\Url}\fi

\bibitem[DeepSeek-AI(2025)]{deepseekv3}
DeepSeek-AI.
\newblock Deepseek-v3 technical report, 2025.
\newblock URL \url{https://arxiv.org/abs/2412.19437}.

\bibitem[Micikevicius et~al.(2018)Micikevicius, Narang, Alben, Diamos, Elsen, Garcia, Ginsburg, Houston, Kuchaiev, Venkatesh, and Wu]{half_training}
Micikevicius, P., Narang, S., Alben, J., Diamos, G., Elsen, E., Garcia, D., Ginsburg, B., Houston, M., Kuchaiev, O., Venkatesh, G., and Wu, H.
\newblock Mixed precision training, 2018.
\newblock URL \url{https://arxiv.org/abs/1710.03740}.

\bibitem[Mitchell et~al.(2023)Mitchell, Tim, Luke, Ari, Ali, and Ludwig]{int8_training}
Mitchell, W., Tim, D., Luke, Z., Ari, M., Ali, F., and Ludwig, S.
\newblock Stable and low-precision training for large-scale vision-language models.
\newblock In Oh, A., Naumann, T., Globerson, A., Saenko, K., Hardt, M., and Levine, S. (eds.), \emph{Advances in {Neural} {Information} {Processing} {Systems}}, volume~36, pp.\  10271--10298. Curran Associates, Inc., 2023.

\bibitem[{NVIDIA}(2022-03-22)]{TMA}
{NVIDIA}.
\newblock Nvidia hopper architecture in-depth.
\newblock \url{https://developer.nvidia.com/blog/nvidia-hopper-architecture-in-depth/}, 2022-03-22.
\newblock Accessed: 2025-06-18.

\bibitem[{NVIDIA Corporation}(2025)]{cudaProgrammingGuide}
{NVIDIA Corporation}.
\newblock Cuda c++ programming guide, 2025.
\newblock URL \url{https://docs.nvidia.com/cuda/cuda-c-programming-guide/index.html#table-alignment-multi-dim-tma}.

\bibitem[Shah et~al.(2024)Shah, Bikshandi, Zhang, Thakkar, Ramani, and Dao]{flashattention3}
Shah, J., Bikshandi, G., Zhang, Y., Thakkar, V., Ramani, P., and Dao, T.
\newblock Flashattention-3: Fast and accurate attention with asynchrony and low-precision, 2024.
\newblock URL \url{https://arxiv.org/abs/2407.08608}.

\bibitem[Zhao et~al.(2025)Zhao, Zhao, Li, and Xu]{deepgemm}
Zhao, C., Zhao, L., Li, J., and Xu, Z.
\newblock Deepgemm: clean and efficient fp8 gemm kernels with fine-grained scaling.
\newblock \url{https://github.com/deepseek-ai/DeepGEMM}, 2025.

\end{thebibliography}
\bibliographystyle{icml2025}

\newpage
\appendix
\onecolumn
\section{Alignment Analysis for Grouped GEMM Operands}
\label{alignment_analysis}
\textbf{Alignment analysis for $A^g$}: 
The tensor $A^g$ participates in Tensor Memory Access (TMA) operations for both global memory loads and shared memory stores. In global memory, the per-row offset of $A^g$ is $K$ bytes, where $K$ denotes the hidden dimension size in Mixture-of-Experts (MoE) architectures. The common practice of setting $K mod 16 = 0$ in modern LLM configurations (Deepseek, Mixtral, Llama) inherently satisfies the 16-byte global memory alignment requirement. 

For shared memory access, we enforce 128-byte alignment during kernel initialization by allocating memory starting from 128-byte aligned addresses. This design makes the alignment condition independent of either the residual row size or $block_K$ parameter.

\textbf{Alignment analysis for $B^g$}: 
The alignment properties of $B^g$ follow similar principles as $A^g$ (detailed analysis omitted for brevity).

\textbf{Alignment analysis for $S^g_A$}:
The scale tensor $S^g_A$ participates in TMA operations involving both global memory loading and shared memory storing. In global memory, the per-row offset is $4\lceil K/128 \rceil$ bytes, which may potentially violate alignment requirements. To address this challenge, we introduce an over-fetching technique described in Section~\ref{subsec:overfetch}. To ensure proper alignment in shared memory, we predefine a 128-byte aligned starting address during kernel configuration.

\textbf{Alignment analysis for $S^g_B$}: 
The tensor $S^g_B$ does not require TMA load operations since each warp group matrix multiply operation typically needs only one or two elements for scaling.

\textbf{Alignment analysis for $C^g$}: 
The output tensor $C^g$ participates in TMA operations for both shared memory loads and global memory stores. In shared memory, the per-row offset is $2 \cdot block_N$ bytes (where the coefficient 2 accounts for half-precision data format). As shown in Section~\ref{subsec:tma_pool}, the second phase of TMA begins at row $res^g - 2^{\lfloor \log_2(res^g) \rfloor}$, which may introduce misalignment. To address this issue, we constrain $block_N$ to be a multiple of 64, ensuring 128-byte alignment regardless of the starting row. For global memory access, the per-row offset is $2N$ bytes. The common practice of setting $2N mod16=0$ in modern LLM architectures (Deepseek, Mixtral, Llama) inherently satisfies the 16-byte alignment requirement.

\section{Example of TMA Runtime Selection and Two-Phase Load-Store}
\label{TMA_example}
\vspace{-0.1in}
\begin{figure}[ht]
\centering
\includegraphics[width=0.8\columnwidth]{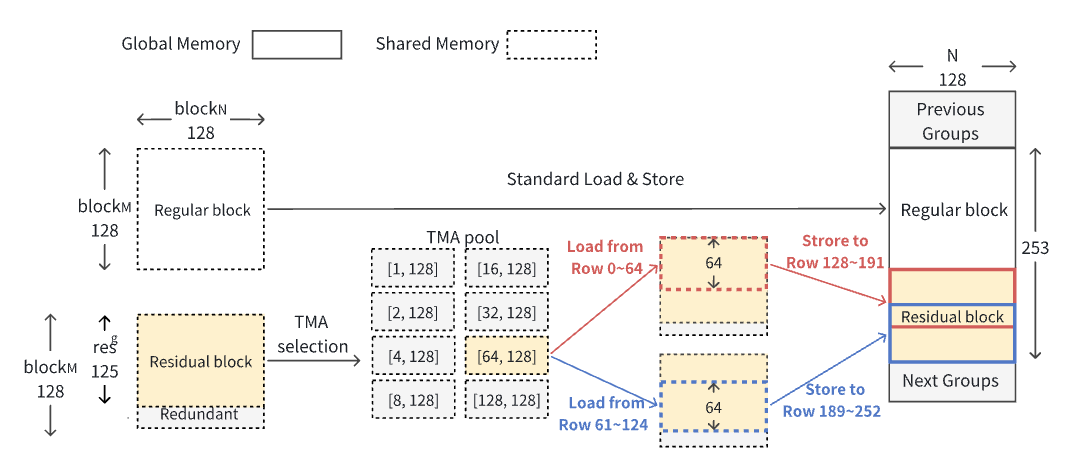}
\vspace{-0.1in}
\caption{An illustration of TMA runtime selection and the two-phase load-store operation. The configuration uses $M^g = 381$ and $N = 128$ for demonstration purposes.}
\label{TMA_load}
\end{figure}

Figure~\ref{TMA_load} demonstrates our method using group $g$ with output matrix dimensions $[M^g, N] = [253, 128]$. With $block\_M = 128$, the first 128 rows (white region) use standard TMA operations with descriptor $[128,128]$, while the residual $res^g = 125$ rows (yellow) require our two-phase approach.

Following Section~\ref{subsec:tma_pool}, we select the optimal descriptor $\mathcal{D}_{opt}^g = [64, 128]$ for the residual portion. The two-phase operation proceeds as:

1) \textbf{First Load-Store}: Transfers rows 0-63 from shared memory to global rows 128-191 (red blocks).

2) \textbf{Second Load-Store}: Loads rows 61-124 from shared memory to global rows 189-252 (blue blocks). The 3-row overlap ensures complete coverage while preventing out-of-bounds writes.

This strategy guarantees that:
\begin{itemize}
\item All 125 residual rows are properly stored
\item No corruption occurs in adjacent memory regions
\item The hardware's static descriptor requirements are maintained
\end{itemize}

\section{Details for experiment}
\subsection{$M^g$ Generation}
\label{M_generation}
To generate the group dimension $M^g$, we employ a randomized algorithm that produces a list of groups $G$ with elements summing to $M$. The generation process consists of the following steps:

\begin{enumerate}
    \item Initialize a zero vector $\mathbf{v}$ of length $G$
    \item For each element $v_i$ in $\mathbf{v}$, assign a random integer value uniformly distributed between $0$ and $2\lfloor M/G \rfloor$
    \item Compute a scaling factor $\alpha = M/\sum_{i=1}^G v_i$ and apply it to each element: $v_i \leftarrow \lfloor \alpha v_i \rfloor$
    \item Adjust the last element $v_G$ to compensate for any residual difference: $v_G \leftarrow v_G + (M - \sum_{i=1}^G v_i)$
\end{enumerate}

This approach ensures that the generated group dimensions maintain the desired total sum $M$ while introducing controlled randomness in the distribution across groups. The scaling operation preserves the relative proportions of the initial random assignments, and the final adjustment guarantees exact sum preservation.

\subsection{Coefficient Matrices}
\begin{table}[htbp]
\centering
\caption{Coefficient matrix of M, N, K, groups and acceleration}
\begin{tabular}{lccccc}
\toprule
 & M & N & K & groups & acceleration (\%) \\
\midrule
M & 1.00 & $4.44\times10^{-16}$ & $7.75\times10^{-17}$ & $-3.43\times10^{-18}$ & 0.0959 \\
N & $4.44\times10^{-16}$ & 1.00 & $-2.50\times10^{-16}$ & $-4.85\times10^{-17}$ & -0.8991 \\
K & $7.75\times10^{-17}$ & $-2.50\times10^{-16}$ & 1.00 & $-2.90\times10^{-17}$ & -0.0274 \\
groups & $-3.43\times10^{-18}$ & $-4.85\times10^{-17}$ & $-2.90\times10^{-17}$ & 1.00 & 0.1036 \\
acceleration (\%) & 0.0959 & -0.8991 & -0.0274 & 0.1036 & 1.00 \\
\bottomrule
\end{tabular}
\end{table}

\begin{table}[htbp]
\centering
\caption{Coefficient matrix of M, N, K, groups and memory saving}
\begin{tabular}{lccccc}
\toprule
 & M & N & K & groups & memory saving (\%) \\
\midrule
M & 1.00 & $4.44\times10^{-16}$ & $7.75\times10^{-17}$ & $-3.43\times10^{-18}$ & -0.5460 \\
N & $4.44\times10^{-16}$ & 1.00 & $-2.50\times10^{-16}$ & $-4.85\times10^{-17}$ & 0.0051 \\
K & $7.75\times10^{-17}$ & $-2.50\times10^{-16}$ & 1.00 & $-2.90\times10^{-17}$ & 0.0036 \\
groups & $-3.43\times10^{-18}$ & $-4.85\times10^{-17}$ & $-2.90\times10^{-17}$ & 1.00 & 0.6356 \\
memory saving (\%) & -0.5460 & 0.0051 & 0.0036 & 0.6356 & 1.00 \\
\bottomrule
\end{tabular}
\end{table}

\end{document}